# X-rays writing/reading of Charge Density Waves in the CuO$_2$ plane of a simple cuprate superconductor

**Gaetano Campi** [1*], **Alessandro Ricci** [2], **Nicola Poccia** [3], **Michela Fratini** [4], **Antonio Bianconi** [1,5]

[1] Institute of Crystallography, CNR, Via Salaria Km 29.300, Monterotondo, Roma, I-00015, Italy.; gaetano.campi@ic.cnr.it

[2] Deutsches Elektronen-Synchrotron DESY, Notkestraße 85, D-22607 Hamburg, Germany; phd.alessandro.ricci@gmail.com

[3] Department of Physics, Harvard University, Cambridge, Massachusetts 02138, USA; npoccia@g.harvard.edu

[4] Fondazione Santa Lucia I.R.C.C.S., Via Ardeatina 306, 00179 Roma, Italy; michela.fratini@gmail.com

[5] Rome International Center of Materials Science (RICMASS), Via dei Sabelli 119A, 00185 Roma, Italy; antonio.bianconi@ricmass.eu

**Abstract:** It is now well established that superconductivity in cuprates competes with charge modulations giving electronic phase separation at a nanoscale. More specifically, superconducting electronic current takes root in the available free space left by electronic charge ordered domains, called charge density wave (CDW) puddles. This means that CDW domain arrangement plays a fundamental role in the mechanism of high temperature superconductivity in cuprates. Here we report about the possibility of controlling the population and spatial organization of the charge density wave puddles in a single crystal La$_2$CuO$_{4+y}$ through X-ray illumination and thermal treatments. We apply a pump-probe method—based on *X-ray illumination* as pump and *X-ray diffraction* as a probe—setting a *writing and reading* procedure of CDW puddles. Our findings are expected to allow new routes for advanced design and manipulation of superconducting pathways in new electronics.

**Keywords:** X ray diffraction; Synchrotron Radiation; Charge Density Waves; High Temperature Superconductivity

## 1. Introduction

The nano-electronics for new generations devices made of complex modern materials is extremely sensitive to changes in the defects disposition [1-3]. To control defects complexity in superconductors some experimental approaches have been proposed in the course of the years. The manipulation of the superconducting grains arrays has been found recently to be possible either by controlling defects self-organization using scanning tips [4] or an external stimulus [5-9]. Recently the improvement of the film deposition techniques have allowed the production of artificial superconducting puddles embedded in a 2D layer intercalated by different block layers in a layered material [10-13]. Although it is common knowledge that a suspended two-dimensional layer should be thermodynamically unstable, and therefore should not exist in nature [14], new advanced materials such as the two-dimensional graphene has recently challenged this paradigm [15-17]. In this material the electronic properties are affected by curved geometries originated by topological defects such as dislocations or ripples.

High temperature superconductors (HTS) are hetero-structures at atomic limit made of intercalated two-dimensional layers. Like graphene, CuO$_2$ layers (with the structure of the 110 plane of a body centered cubic, bcc, lattice) in cuprates [18,19], as well as B layers in diborides [20-22] or FeSe layers in A$_x$Fe$_{2-y}$Se$_2$ chalcogenides [23] are truly two-dimensional crystal (just one atom thick). Because of the general instability with respect to bending fluctuations, these conductive planes are rippled and these effects are enormously relevant on their structural, thermodynamic and electronic properties. Using strain engineering it is possible to modify in a desirable way the electronic structure of atomic layers with exciting perspectives for electronics. In case of HTS, electronic organization in modulated structures called charge density waves, CDW, have been found to play an essential role in superconducting mechanism [24-27]. In fact, it has been established that electronic superconducting currents develop in the space free from CDW [28,29]. In this framework, the possibility to control both the population and the spatial arrangement of the CDW puddles becomes a key point to be investigated for the different possible applications.

CDW in cuprates have been found recently through the real space visualization of the 3D superlattices measured by scanning micro X-ray diffraction in different single crystals [8,9,18,23,28-30]. These transverse modulations are similar to those found in the superconducting Bi2212, characterized by the wave-vectors 0.21$b$* [31-32]. They are due to stripes with period 4.8b along the diagonal direction $b$. It has been found that the CDW puddles are spatially anti-correlated with other spin and charge ordered puddles and with domains rich of interstitial dopants oxygen defects [28,33-36], in a scenario of frustrated phase separation which is a common feature of doped cuprates [37-41]. Either oxygen interstitials rich domains or CDW puddles are sensitive to X-ray illumination in a different temperature range [42-43] but together work to establish the optimum inhomogeneity, which raises the critical temperature to the optimum value [18]. Here we study CDW puddles in the oxygen doped orthorhombic La$_2$CuO$_{4+y}$ (LCO) superconductor showing the details of the manipulation process, which allow the tuning the order of the CDW puddles in the LCO system.

## 2. Results

2.1 Temperature dependence of CDW puddles.

Indexing of superlattice peaks around the Bragg lattice reflections show the presence of incommensurate CDW modulation with wave-vector **q**$_{CDW}$=(0.023$a$* 0.21$b$* 0.29$c$*) where $a$*, $b$* and $c$* are the unit vectors of the reciprocal lattice. The profiles of this superstructure peak, measured at 300K and 100K in the $a^*b^*$ plane and fitted by Gaussian line shape, are shown in Figure 1a and 1b.

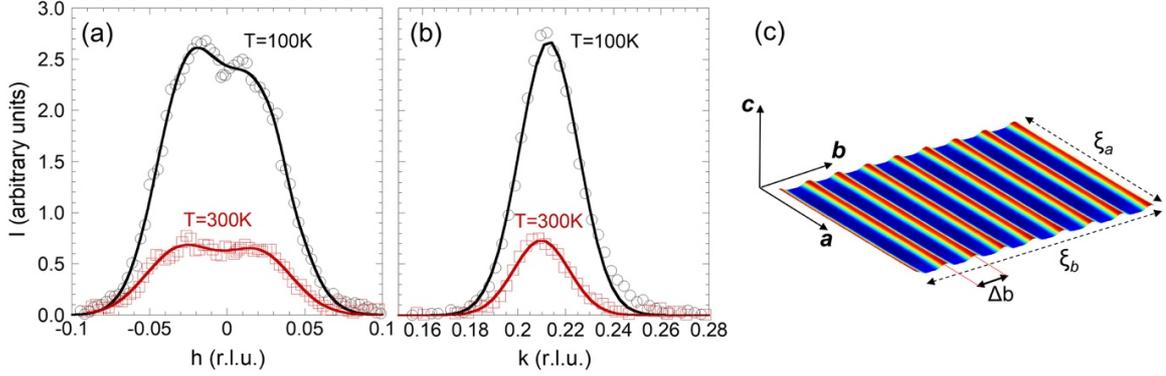

**Figure 1.** Profiles of the $q_{CDW}$ superstructure along the (a) $a^*$ and (b) $b^*$ directions. The profiles, measured around the 006 Bragg peak have been fitted by Gaussian lineshape (continuous lines). The peaks along the **a*** direction show quite diffuse scattering covering a larger h-range around zero. A typical single CDW puddle in $La_2CuO_{4+y}$ in the ab plane with size of $\xi_a \times \xi_b$. $\Delta b$ represents the period of the CDW, corresponding to 4.76 unit cells along the crystal b axis.

The modulations along $a^*$ give X-ray diffuse superlattice peaks due to a quasi one-dimensional (1D) modulation associated with stripes running in $a$ direction with a period of about 2.6 nm in the $b$ direction. These incommensurate Charge Density Wave (CDW), are established in nanoscale regions, as results by the analysis of the widths of X-ray satellites. Indeed, the FWHM of the CDW peak, along the $a^*$ and along the $b^*$ direction are related with the correlation length of the CDW puddles $\xi_a = a/FWHM(a^*)$ and $\xi_b = b/FWHM(b^*)$. We have found 11 ± 2 nm and 19 ± 2 nm for $\xi_a$ and $\xi_b$, respectively. A pictorial view of the $q_{CDW}$ modulation is given in Figure 1c, where $a$, $b$ and $c$ represent the crystalline directions of the orthorhombic lattice.

The temperature dependence of the CDW order is shown in Figure 2. The normalized profiles along the $a^*$ and $b^*$ directions are visualized by upper and lower panel in Figure 2a, respectively. The integrated intensity of the CDW peak increases by cooling the sample from room temperature down to 100 K, as shown in Figure 2b. We find a CDW onset temperature of 250 K.

2.2 Controlling CDW puddles by X ray illumination.

The possibility to photo generate CDW puddles in $La_2CuO_{4+y}$ has been previously shown in ref. [18]. Our pump and probe experimental approach, based on X-ray illumination and diffraction has been performed with synchrotron X-ray photons focused onto a spot size of 100 μm² on the sample surface.

We have found that the effect of photo generation of CDW in the x-ray energy range between 8 KeV and 12KeV does not depend on photon energy but on the x-ray illumination fluence. Further new measurements are required to investigate the pump and probe process in the soft X-ray and optical regimes. The photon flux (defined as the number of photons hitting the sample surface per second per unit area) is given by $\Phi_{P(0.1nm)} = 5 \times 10^{14} N_{P(0.1nm)} s^{-1} cm^{-2}$ and corresponds to the power density of $1\ W\ cm^{-2}$. In our method the X-ray photon beam works at the same time as pump and as probe on a surface layer with an area of 100 μm² and about 1.5 μm thick. In this volume the physical state of the system is controlled by the fluence $F_{P(0.1nm)}(N_{P(0.1nm)} \cdot cm^{-2}) = \Phi_P \cdot t$.

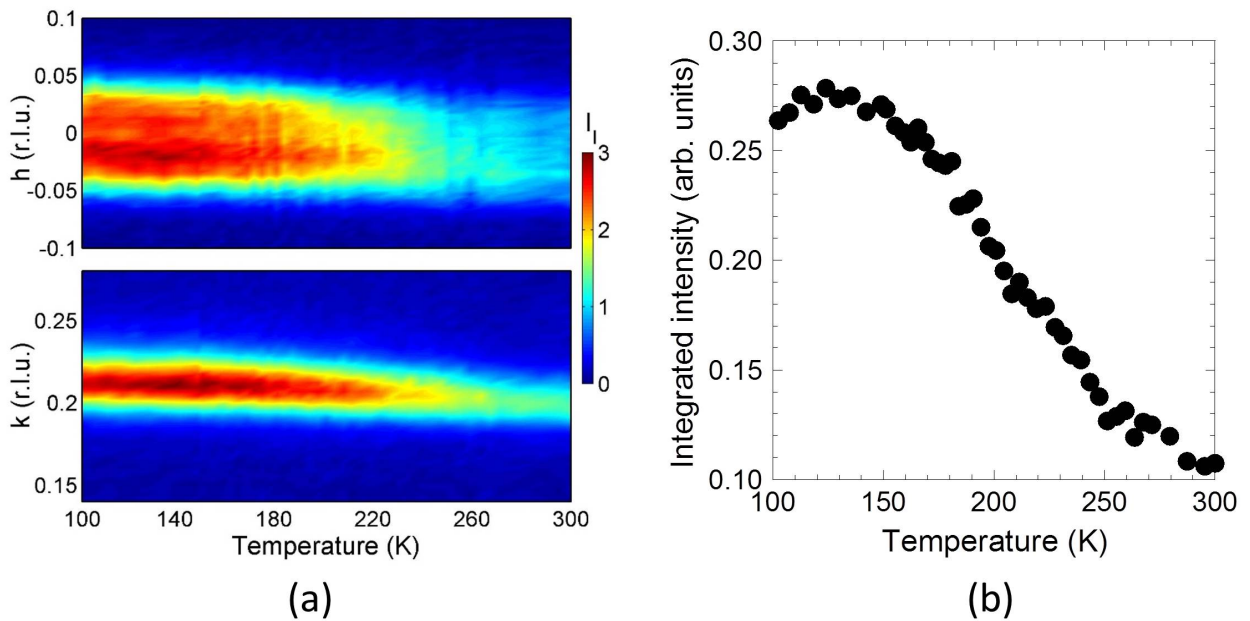

**Figure 2.** (a) Map of normalized diffraction profiles q$_{CDW}$ along (upper panel) ***a*** and (lower panel) ***b**** direction during cooling from room temperature down to T=100 K. (b) Integrated intensity of the q$_{CDW}$ peak during the cooling.

We quenched the sample from room temperature, where CDW puddles population is weak, down to 100K, leading the system to a frozen metastable disordered state. Than we illuminated continuously the sample. The pump and probe results are illustrated in Figure 3.

The color maps obtained from the normalized profiles, of $q_{CDW}(a^*)$ and $q_{CDW}(b^*)$ shown in the upper and lower panels of Figure 3a, respectively, visualize the evolution of CDW during the X-ray illumination. The CDW peak integrated intensity increases of about three times under illumination, as shown by the integrated intensity of $q_{CDW}$ peak in Figure 3B.

A single oscillation in each puddle is given by $\gamma(b) = b/q_{CDW}(b^*) = 2.5$ nm and $\gamma(a) = a/q_{CDW}(a^*) = 23.4$ nm along the **b** and **a** direction, respectively. Thus, each puddle contains a number of oscillations given by $\xi(a)/\gamma(a)=0.5$ and $\xi(b)/\gamma(b)=7$ in the a and b directions, respectively. Thus we have $\xi(a)$ x $\xi(b)$ puddles made of about 7 unidirectional stripes each. A pictorial view of CDW puddles is shown in the upper part of Figure 4.

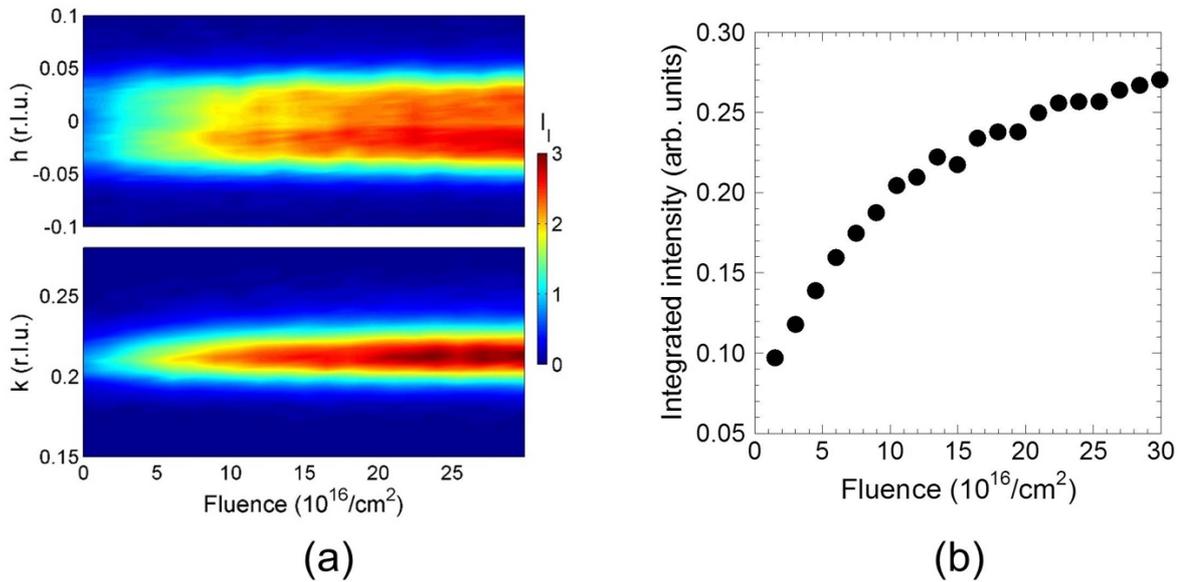

**Figure 3.** (a) Map of normalized diffraction profiles $q_{CDW}$ along (upper panel) $a^*$ and (lower panel) $b^*$ direction during the X-ray continuous illumination at T=100 K. (b) Integrated intensity of the $q_{CDW}$ peak increases by a factor about 3 during the illumination.

2.3 CDW Writing/reading procedure.

Thus, we have shown the feasibility of tuning the CDW population through X-ray illumination and also the possibility to erase such order by thermal treatments. These are the elementary steps for a writing/reading process. We have performed therefore a series of experiment of direct manipulation of the CDW puddles in real space in a La$_2$CuO$_{4+y}$ sample through local X-ray continuous illumination at low temperatures.

Figure 4a shows the color-map of the intensity of the CDW peak, along $b^*$, obtained by scanning the sample using an X-ray beam of 100 μm$^2$ and a low fluence of 1.5 (10$^{16}$/cm$^2$). The histogram in Figure 4c shows a quite flat distribution of the CDW intensity, but not zero. In fact, in order to have a successful X-ray illumination growth of the CDW puddles, the region irradiated must already have some seeds of CDW puddles, as shown by the light blue stripe in the map centre of Figure 4a.

We irradiated three spots indicated by 1, 2, 3 using a X-rays beam of 100μm² for a total fluence of 50 (10$^{16}$/cm²); the effects of this continuous local illumination are than monitored by re-scanning the sample using the low fluence of 1.5 (10$^{16}$/cm²). The results are shown in the map of Figure 4b by the three bright spots at the positions 1, 2, 3 of increased integral intensity, respectively shown also in the histogram of Figure 4d.

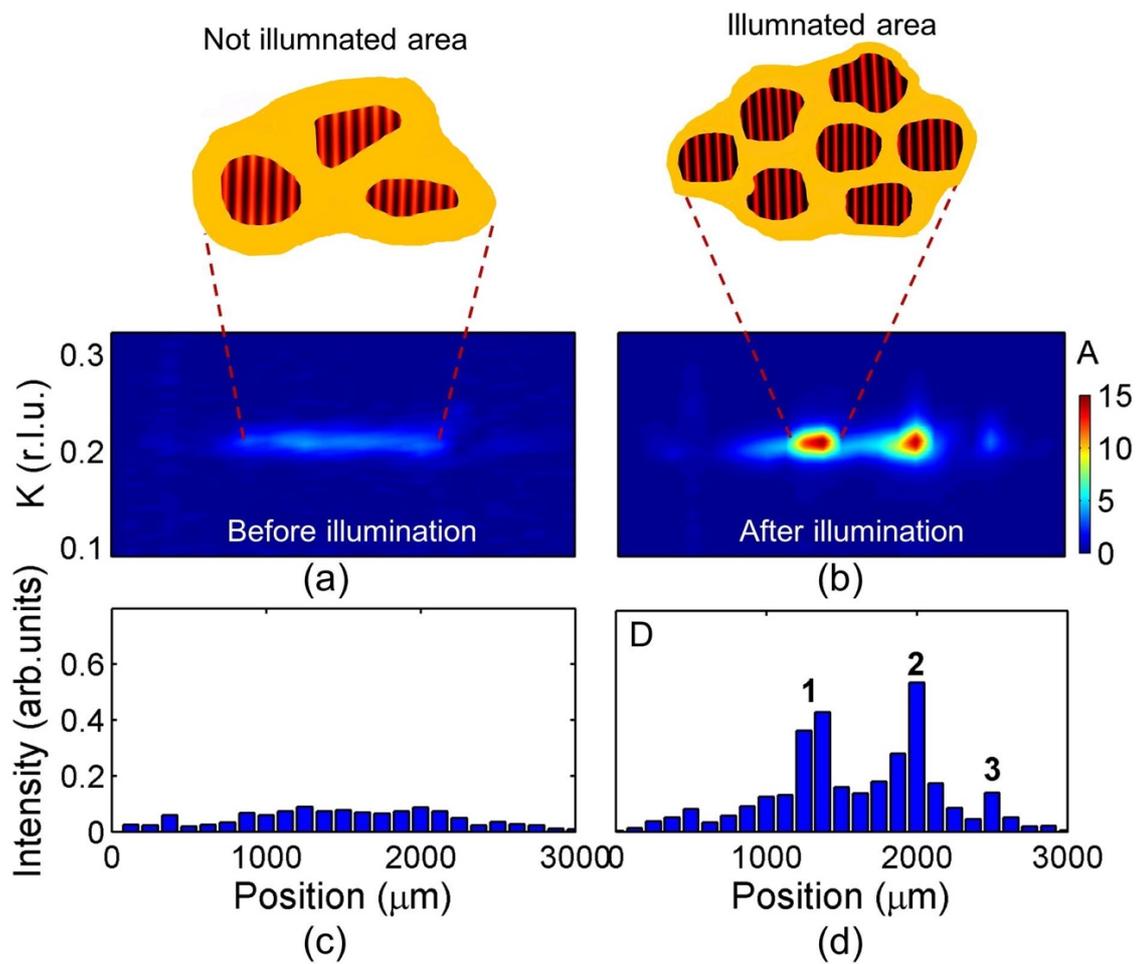

**Figure 4.** Writing/reading local CDW rich puddles domains by X-ray illumination. Diffraction scanning profiles along the sample (a) before and (b) after X-ray illumination by 100 microns size beam in the points indicated with 1, 2, 3. Scanning profile integral area (c) before and (d) after X-ray illumination.

## 3. Discussion

The relevance of the possibility to control and manipulate the spatial organization of CDW puddles populations lies in the fact that the free electrons form superconducting currents flowing in the interstitial space [28,29]. Thus, changing "*point by point*" the population of the CDW puddles leads to modify the interstitial space available for the electron currents. In atomic layers intrinsic thermodynamic instabilities give rise to charge (spin, orbital) ordering phenomena. Easy method such as thermal cycling and X ray continuous illumination with focused beams allows to write/reading and erasing arbitrary CDW pathways. This will allow to design new devices with controlled functionality.

## 4. Materials and Methods

Diffraction measurements on a single crystal of underdoped La$_2$CuO$_{4+y}$ of size 3x2x0.5 mm$^3$, were performed on the crystallography beamline, XRD1, at ELETTRA [44]. The sample was grown first as La$_2$CuO$_4$ by flux method and then doped by electrochemical oxidation. The X-ray beam, emitted by the wiggler source of 2 GeV electron storage ring, was monochromatized by a Si(111) double crystal, and focused on the sample. The superconducting transitions of samples have been established using contactless single-coil inductance as described elsewhere [5,16,30]. The temperature of the crystal was monitored with an accuracy of ±1K. We have collected the data in the reflection geometry using a CCD detector, a photon energy of 12.4 KeV (wavelength $\lambda$=1Å). The sample oscillation around the b-axis was in a range 0<Φ<20°, where Φ is the angle between the direction of the photon beam and the a-axis. We have investigated a portion of the reciprocal space up to 0.6 Å$^{-1}$ momentum transfer, i.e., recording the diffraction spots up to the maximum indexes 3, 3, 19 in the *a\**, *b\**, *c\** direction respectively. Thanks to the high brilliance source, it has been possible to record a large number of weak superstructure spots around the main peaks of the average structure. Twinning of the crystal has been taken into account to index the superstructure peaks. The orthorhombic lattice parameters of single crystal were determined to be a=(5.345±0.008)Å, b=(5.386±0.004)Å, c=(13.205±0.031)Å at room temperature. The space group of the sample is *Fmmm*.

## 5. Conclusions

In conclusion in this paper we report the study of CDW in an orthorhombic crystal La$_2$CuO$_{4+y}$ where dopants are mobile oxygen interstitials which has unique features in comparison with CDW in different cuprate families (45-48): the CDW wave-vector has been found to be aligned in the long b orthorhombic axis in the diagonal direction from the Cu-O-Cu direction like in nikelates, cobaltates and in doped La124 family at very low doping. Moreover the average CDW puddles are formed by only 7 oscillations which is larger than in Hg1201 but smaller than in other systems. Moreover the photo-stimulated CDW puddles in La$_2$CuO$_{4+y}$ are aligned in the b-axis direction and form a smectic phase in the same crystal while in underdoped YBa$_2$Cu$_3$O$_{6+y}$ and Hg1201 form a nematic phase. These results support the increasing accumulating experimental evidence that the CDW character in cuprates changes in different families (45-48).

In summary we have shown the details of the manipulation process which allow to tune the order of the CDW puddles in the LCO system. This approach has unveiled the condition and the process of formation of Charge Density Waves puddles in the CuO$_2$ plane and shows a possible method for their direct control. In particular we have shown that the synthesis of photo-stimulated CDW puddles in the LCO can be controlled by setting the fluence of synchrotron x-ray illumination and thermal cycling. Thanks to the advanced X ray optics it will be possible to write and read CDW pathways whose size is determined by the X ray beam size. This new approach in arbitrary nanostructures manipulation promises increasingly large scale integration of conventional devices.

**Acknowledgments:** The authors thank Luisa Barba and XRD1 beamline staff at ELETTRA, Trieste, Italy.

**Author Contributions:** G.C. and A.B. conceived and designed the experiments; G.C. and M.F. performed the experiments; G.C., A.R. and M.F. analyzed the data; G.C., N.P. and A.B. wrote the paper.

**Conflicts of Interest:** The authors declare no conflict of interest.